\documentclass[english,twoside,12pt]{amsart}
\usepackage[T1]{fontenc}
\usepackage{amsthm}
\usepackage[mac]{inputenc}
\usepackage{amsfonts,amsmath,epsfig,amscd,amssymb,latexsym}
\theoremstyle{plain}
\newtheorem{theorem}{Theorem}

\newtheorem{corollary}[theorem]{Corollary}

\theoremstyle{definition}

\title{A short note on the kissing number of the lattice in Gaussian wiretap coding}
\author{Anne-Maria Ernvall-Hyt\"onen}\address{Department of Mathematics and Statistics, University of Helsinki, Finland}
\thanks{The author would like to thank Prof. Frederique Oggier and Fuchun Lin for an inspiring conversation.}
\begin{document}
\maketitle
\begin{abstract}
We show that on an $n=24m+8k$-dimensional even unimodular lattice, if the shortest vector length is $\geq 2m$, then as the number of vectors of length $2m$ decreases, the secrecy gain increases. We will also prove a similar result on general unimodular lattices. Furthermore, assuming the conjecture by Belfiore and Sol\'e, we will calculate the difference between inverses of secrecy gains as the number of vectors varies. Finally, we will show by an example that there exist two lattices in the same dimension with the same shortest vector length and the same kissing number, but different secrecy gains.
\end{abstract}
\section{Introduction}
Belfiore and Oggier defined in \cite{belfioggiswire} the secrecy gain
\[
\chi_{\Lambda}=\max_{y\in \mathbb{R}, 0<y}\frac{\Theta_{\mathbb{Z}^n}(yi)}{\Theta_{\Lambda}(yi)},
\]
where
\[
\Theta_{\Lambda}(z)=\sum_{x\in \Lambda}e^{\pi i ||x||^2z}
\]
as a new lattice invariant to measure how much confusion the eavesdropper will experience while the lattice $\Lambda$ is used in Gaussian wiretap coding. The function $\Xi_{\Lambda}(y)=\frac{\Theta_{\mathbb{Z}^n}(yi)}{\Theta_{\Lambda}(yi)}$ is called the secrecy function. Belfiore and Sol\'e then conjectured in \cite{belfisole} that the secrecy function attains its maximum at $y=1$, which would then be the value of the secrecy gain. Ernvall-Hyt\"onen \cite{oma:ieee} proved this all known (and for some possibly existing) extremal lattices), and derived a method to prove or disprove the conjecture for any given unimodular lattice.The secrecy gain was further studied by Oggier, Sol\'e and Belfiore in \cite{belfisoleoggis} and by Lin and Oggier in \cite{oglinitw}.

Recently, Lin and Oggier considered unimodular lattices in dimensions $8< n\leq 23$ \cite{oggierlin}, and furthermore, they considered the dependence of the secrecy gain on the kissing number $K(\Lambda)$ of the lattice. They proved that in dimensions $16\leq n\leq 23$, for non-extremal lattices the secrecy gain is given by
\[
\chi_{\Lambda}=\frac{1}{1-\frac{2n}{2^6}+\frac{2n(n-23)+K(\Lambda)}{2^{12}}}.
\]
In particular, in this case, they proved that the smaller the kissing number, the better the secrecy gain.

The question whether one can use kissing number to find the best secrecy gain 
in general, is not that straightforward to answer: In some cases, the kissing number determines the secrecy gain, but not always.

We prove that if an even unimodular lattice in dimension $n=24m+8k$ ($k\in \{0,1,2\}$) has the shortest vector length $\geq 2m$, then the secrecy gain increases as the number of vectors of length $2m$ decreases (ie. when the kissing number decreases, or as a limit case: when there are no vectors of length $2m$ but the shortest vector length is $2m+2$, i.e. the lattice is extremal). In particular, this shows that the secrecy gain is better on extremal lattices than on lattices with vectors of length $2m$. We will also prove a similar theorem on odd lattices: if all vectors are of length at least $\left\lfloor \frac{n}{8}\right\rfloor$, then the secrecy gain increases as the number of vectors of length $\left\lfloor \frac{n}{8}\right\rfloor$ decreases. It would be possible to prove that if all the vectors are of length $\geq\left\lfloor \frac{n}{8}\right\rfloor+1$, then the secrecy gain is better than if the shortest vector length is $\left\lfloor \frac{n}{8}\right\rfloor$. However, since all the lattices with shortest vector length $\left\lfloor \frac{n}{8}\right\rfloor+1$ are known, this would not give any new information past the comparisons in article \cite{oggierlin} (see \cite{conwayodlyzkosloane} for why there are very few lattices like this).
\section{Preliminaries}
For information on theta function, one can study the book \cite{steinshakarchi} by Stein and Shakarchi. For an extensive source on lattices, one may be referred to the book \cite{conwaysloane} by Conway and Sloane. However, to increase the readability of the current article, we will briefly recall the basic facts.

Define first the following theta functions:
\begin{alignat*}{1}
\vartheta_2(\tau) & =e^{\pi i \tau/4}\prod_{n=1}^{\infty}(1-q^{2n})(1+q^{2n})(1+q^{2n-2})\\
\vartheta_3(\tau) &=\prod_{n=1}^{\infty}(1-q^{2n})(1+q^{2n-1})^2\\
\vartheta_4(\tau) &  =\prod_{n=1}^{\infty}(1-q^{2n})(1-q^{2n-1})^2.
\end{alignat*}

Notice that $\vartheta_3$ is the theta function of the lattice $\mathbb{Z}^n$.
A lattice $\Lambda$ is called unimodular if its determinant $=\pm 1$, and the norms are integral, ie, $||x||^2\in \mathbb{Z}$ for all vectors $x\in \Lambda$. Further, it is called even, if $||x||^2$ is even for all $x\in\Lambda$. Otherwise it is called odd. A lattice can be even unimodular only if the dimension is divisible by $8$. Odd unimodular lattices exist in all dimensions: $\mathbb{Z}^n$ is an example of such.

Theta functions of even unimodular lattices in dimension $n=24m+8k$ ($k\in \{0,1,2\}$) can be written as polynomials
\[
\Theta=E_4^{3m+k}+\sum_{j=1}^m b_j E_3^{3(m-j)+k}\Delta^j,
\]
where $E_4=\frac{1}{2}\left(\vartheta_2^8+\vartheta_3^8+\vartheta_4^8\right)$ and $\Delta=\frac{1}{256}\vartheta_2^8\vartheta_3^8\vartheta_4^8$. Here $E_4$ is an Eisenstein series, and $\Delta$ a discriminant function.

Generally, theta functions of unimodular lattices in dimension $n=8\mu+\nu$ ($\nu\in \{0,1,\dots,7\}$) can be written as polynomials:
\[
\Theta_{\Lambda}=\sum_{r=0}^{\mu}a_r\vartheta_3^{n-8r}\Delta_8^r,
\]
where $\Delta_8=\frac{1}{16}\vartheta_2^4\vartheta_4^4$. Notice that this gives an alternative representation for theta functions of even lattices.

We call an even lattice \emph{extremal}, if the shortest vectors are of length $2m+2$. Earlier, the definition of an extremal lattice stated that a lattice is extremal if the shortest vectors are of length $\left\lfloor\frac{n}{8}\right\rfloor+1$. However, Conway, Odlyzko and Sloane were able to show that there are very few extremal lattices with this definition, and they all exist in dimensions $\leq 24$ \cite{conwayodlyzkosloane}.

\section{Increasing the secrecy gain by decreasing the number of short vectors}
In this section, we will prove two theorems, first of which corresponds to even lattices, and the second one to all unimodular lattices.
\begin{theorem}\label{melkein} Let $\Lambda$ be an even unimodular lattice in the dimension $n=24m+8k$ with $k\in \{0,1,2\}$ with the shortest vector length $\geq 2m$. Let $k_{2m}\geq 0$ be the number of vectors of the length $2m$. Then the secrecy gain increases as $k_{2m}$ decreases. Assuming the conjecture by Belfiore and Sol\'e, the difference between the inverses of the secrecy gains of two lattices with $k_{2m}$ and $k_{2m}'$ vectors of length $2m$, is $(k_{2m}-k_{2m}')\frac{3^k}{4^{6m+k}}$.
\end{theorem}
By letting $k_{2m}=0$ in the previous theorem, we have the following special case:
\begin{corollary}
The secrecy gain of an $n$-dimensional extremal even unimodular lattice, when $n=24m+8k$ ($k\in \{0,1,2\}$), is better than the secrecy gain of any even $n$-dimensional unimodular lattice with shortest vector length $2m$.
\end{corollary}
Let us now move to the proof of Theorem \ref{melkein}:
\begin{proof}
the theta series of the lattice can be written as a polynomial of the Eisenstein series $E_4$ and the discriminant function $\Delta$:
\[
\Theta=E_4^{3m+k}+\sum_{j=1}^m b_j E^{3(m-j)+k}\Delta^j.
\]
Ernvall-Hyt\"onen showed in \cite{oma:ieee} that the  secrecy gain is the maximal value of
\[
\left((1-z)^{3m+k}+\sum_{j=1}^m \frac{b_j}{256^j}(1-z)^{3(m-j)+k}z^{2j}\right)^{-1}
\]
in the range $z\in \left(0,\frac{1}{4}\right]$. Now we need to find how this expression changes when the kissing number changes.
Write
\[
E_4^h\Delta^j=q^{2j}+a_{h,j,1}q^{2j+2}+a_{h,j,2}q^{2j+4}+\cdots,
\]
where $q=e^{\pi i}$. If the theta function of an even unimodular $24m+8k$-dimensional ($k\in \{0,1,2\}$) lattice is of the form
\[
1+k_{2m}q^{2m}+\cdots,
\]
then to derive the coefficients $b_j$, we have to solve the following system of equations:
\[
\left\{\begin{array}{r} a_{3m+k,0,1}+b_1=0 \\
a_{3m+k,0,2}+b_1a_{3(m-1)+k,1,1}+b_2=0  \\
a_{3m+k,0,3}+b_1a_{3(m-1)+k,1,2}+b_2a_{3(m-2),2,1}+b_3=0  \\
\cdots \quad \quad \quad \cdots \\
a_{3m+k,0,m-1}+b_1a_{3(m-1)+k,1,m-2}+\cdots +b_{m-2}a_{3+k,m-2,1}+b_{m-1}=0\\
a_{3m+k,0,m}+b_1a_{3(m-1)+k,1,m-1}+\cdots +b_{m-1}a_{k,m-1,1}+b_m=k_{2m} 
\end{array}\right.
\]
From this system of equations, it is clear that $b_1,b_2,\cdots ,b_{m-1}$ depend only on the coefficients $a_{i,j,h}$ for suitable values of $(i,j,h)$, and not on the number $k_{2m}$. We may write
\[
b_m=k_{2m}-\left(a_{3m+k,0,m}+b_1a_{3(m-1)+k,1,m-1}+\cdots +b_{m-1}a_{k,m-1,1}\right).
\]
Let us now compare the secrecy gains of lattices $\Lambda$ and $\Lambda'$. Assume $k_{2m}>k_{2m}'$.  Denote the corresponding coefficients $b_j$ and $b_j'$, respectively. Then $b_j=b_j'$,when $j<m$, and
\begin{multline*}b_m-b_m'=k_{2m}-\left(a_{3m+k,0,m}+b_1a_{3(m-1)+k,1,m-1}+\cdots +b_{m-1}a_{k,m-1,1}\right)\\-\left(k_{2m'}-\left(a_{3m+k,0,m}+b_1a_{3(m-1)+k,1,m-1}+\cdots +b_{m-1}a_{k,m-1,1}\right)\right)\\=k_{2m}-k_{2m}'>0.
\end{multline*}
 Now the secrecy function of the lattice $\Lambda$ can be written and estimated in the following way:
\begin{multline*}
\Xi_{\Lambda}(y)=\frac{\vartheta_3^n(y)}{\Theta_{\Lambda}(y)}\\=\left(\left(1-\frac{\vartheta_2^4\vartheta_4^4}{\vartheta_3^8}(y)\right)^{3m+k}\right.\left.+\sum_{j=1}^m\frac{b_j}{256^j}\left(1-\frac{\vartheta_2^4\vartheta_4^4}{\vartheta_3^8}(y)\right)^{3(m-j)+k}\cdot\left(\frac{\vartheta_2^4\vartheta_4^4}{\vartheta_3^8}(y)\right)^{2j}\right)^{-1}\\<\left(\left(1-\frac{\vartheta_2^4\vartheta_4^4}{\vartheta_3^8}(y)\right)^{3m+k}+\sum_{j=1}^m\frac{b_j'}{256^j}\left(1-\frac{\vartheta_2^4\vartheta_4^4}{\vartheta_3^8}(y)\right)^{3(m-j)+k}\cdot\left(\frac{\vartheta_2^4\vartheta_4^4}{\vartheta_3^8}(y)\right)^{2j}\right)^{-1}\\=\Xi_{\Lambda'}(y),
\end{multline*}
which proves the first claim.

Let us now move to the proof of the second claim. Assuming that the secrecy gain conjecture holds, the secrecy gain is obtained at $y=1$. Recall $\frac{\vartheta_2^4\vartheta_4^4}{\vartheta_3^8}(1)=\frac{1}{4}$. Now we just need to calculate the difference $\Xi_{\Lambda}^{-1}(1)-\Xi_{\Lambda'}^{-1}(1)$:
\begin{multline*}
\Xi_{\Lambda}^{-1}(1)-\Xi_{\Lambda'}^{-1}(1)\\=\left(\left(1-\frac{\vartheta_2^4\vartheta_4^4}{\vartheta_3^8}(1)\right)^{3m+k}\right.\left.+\sum_{j=1}^m\frac{b_j}{256^j}\left(1-\frac{\vartheta_2^4\vartheta_4^4}{\vartheta_3^8}(1)\right)^{3(m-j)+k}\cdot\left(\frac{\vartheta_2^4\vartheta_4^4}{\vartheta_3^8}(1)\right)^{2j}\right)\\-\left(\left(1-\frac{\vartheta_2^4\vartheta_4^4}{\vartheta_3^8}(1)\right)^{3m+k}+\sum_{j=1}^m\frac{b_j'}{256^j}\left(1-\frac{\vartheta_2^4\vartheta_4^4}{\vartheta_3^8}(1)\right)^{3(m-j)+k}\cdot\left(\frac{\vartheta_2^4\vartheta_4^4}{\vartheta_3^8}(1)\right)^{2j}\right)\\=\frac{b_m-b_m'}{256^m}\left(1-\frac{\vartheta_2^4\vartheta_4^4}{\vartheta_3^8}(1)\right)^{k}\cdot\left(\frac{\vartheta_2^4\vartheta_4^4}{\vartheta_3^8}(1)\right)^{2m}=(b_m-b_m')\frac{3^{2m}}{4^{6m+k}},\end{multline*}
and now the proof is complete.
\end{proof}

One may also prove the following theorem:
\begin{theorem}\label{melkein_pariton} Let $\Lambda$ be an odd unimodular lattice in the dimension $n$ with the shortest vector length $\left\lfloor \frac{n}{8}\right\rfloor$. Let the number of vectors of length $\left\lfloor \frac{n}{8}\right\rfloor$ be $h$. Then, when $h$ decreases, the secrecy gain increases. Furthermore, assuming the conjecture by Belfiore and Sol\'e, if lattices $\Lambda$ and $\Lambda'$ have $h$ and $h'$ vectors of length $\left\lfloor \frac{n}{8}\right\rfloor$, respectively, then the difference of the inverses of the secrecy gains is $\frac{h-h'}{4^{5\left\lfloor \frac{n}{8}\right\rfloor}}.$
\end{theorem}
Furthermore, one could formulate a similar corollary as in the case of even lattices. However, since all the lattices with shortest vector length $\left\lfloor \frac{n}{8}\right\rfloor +1$ are known, this doesn't give any new information compared to \cite{oggierlin}.

\begin{proof}
Since the proof of Theorem \ref{melkein_pariton} is similar to the proof of Theorem \ref{melkein}, we will only sketch the proof to point the differences.

Since the theta function of any unimodular lattice has the polynomial representation
\[
\Theta_{\Lambda}=\sum_{r=0}^{\left\lfloor \frac{n}{8}\right\rfloor}a_r\vartheta_3^{n-8r}\Delta_8^r,
\]
writing
\[
\vartheta_3^r\Delta_8^s=q^s+c_{r,s,1}q^{s+1}+c_{r,s,2}q^{s+2}+\cdots,
\]
we get the system of equations
\[
\left\{\begin{array}{r}a_0=1\\ a_0c_{n,0,1}+a_1=0\\\cdots\quad\cdots\\a_0c_{n,0,\left\lfloor\frac{n}{8}\right\rfloor}+a_1c_{n-8,1,\left\lfloor\frac{n}{8}\right\rfloor-1}+\cdots+a_{\left\lfloor\frac{n}{8}\right\rfloor}=h.\end{array}\right.
\]
We may now proceed just like in the proof of the previous theorem. To prove the second part of the theorem, we work just like in the proof of the previous theorem. We use the polynomial expression for the inverse of the secrecy function:
\[
\Xi_{\Lambda}^{-1}=\sum_{r=0}^{\mu}\frac{a_r}{16^r}\frac{\vartheta_2^{4r}\vartheta_4^{4r}}{\vartheta_3^{8r}}
\]
and notice that in the difference between the inverses, only the last terms in the polynomials remain. Their difference is
\[
\frac{(h-h')}{16^{\left\lfloor\frac{n}{8}\right\rfloor}}\cdot \left(\frac{\vartheta_2^{4}\vartheta_4^{4}}{\vartheta_3^{8}}\right)^{\left\lfloor\frac{n}{8}\right\rfloor}.
\]
If the conjecture by Belfiore and Sol\'e holds, then the maximum is obtained at $y=1$, i.e. $\frac{\vartheta_2^{4}\vartheta_4^{4}}{\vartheta_3^{8}}=\frac{1}{4}$, then the difference becomes
\[
\frac{h-h'}{4^{5\left\lfloor\frac{n}{8}\right\rfloor}},
\]
which completes the proof.
\end{proof}

\section{Same kissing numbers, different secrecy gains}
We will show that the extremal even unimodular lattice in dimension $40$ with shortest vector length $4$ and kissing number $39600$ has a different secrecy gain than the odd unimodular lattice in dimension $40$ with the same shortest vector length and kissing number. This lattice is also extremal, in the sense that it has the longest shortest vectors.

It was showed in \cite{oma:ieee} that the $40$-dimensional even unimodular lattice satisfies the secrecy gain conjecture. Let us use the methods from there to find the actual value of the secrecy gain.  The extremal even unimodular lattices in dimension $40$ have theta series are of the form
\[
1+39600q^4+\cdots.
\]
The theta function of the lattice can also be written as
\[
E_4^5-1200E_4^2\Delta,
\]
and therefore, the secrecy gain is the maximal value of the function $\left((1-z)^5-\frac{75}{16}z^2(1-z)^2\right)^{-1}$ on the interval $\left(0,\frac{1}{4}\right]$, which is obtained at $z=\frac{1}{4}$, and this value is $\frac{4096}{297}$.

Consider now the odd unimodular lattices in dimension $40$ with theta series of the form
\[
1+39600q^4+1048576q^5+\cdots.
\]
More on these odd lattices can be found at \cite{harada}. Comparing coefficients, we see that if the theta function of a unimodular lattice is of the form
\[
1+39600q^4+1048576q^5,
\]
then the theta function can be represented as (see \cite{harada})
\[
\vartheta_3^{40}-80\vartheta_3^{32}\Delta_8+1360\vartheta_3^{24}\Delta_8^2-2560\vartheta_3^{16}\Delta_8^3+20480\vartheta_3^8\Delta_8^4.
\]
Using the method from \cite{oma:ieee}, we'll see that it is sufficient to show that the polynomial
\[
1-5z+\frac{1360}{16^2}z^2-\frac{2560}{16^3}z^3+\frac{20480}{16^4}z^4
\]
obtains its minimal value on the interval $\left(0,\frac{1}{4}\right]$ at $z=\frac{1}{4}$, and then to compute the inverse of this value.

To show that the polynomial obtains its minimum at $z=\frac{1}{4}$, let us first differentiate it. The derivative is
\[
\frac{5}{8}\left(2z^3-3z^2+17z-8\right)\leq \frac{5}{8}\left(2z^3+17z-8\right)\leq \frac{5}{8}\left(\frac{2}{64}+\frac{17}{4}-8\right)<0
\]
on the interval $z\in \left(0,\frac{1}{4}\right]$. Hence, the polynomial obtains its minimum at $z=\frac{1}{4}$, and this value is $\frac{301}{4096}$. Hence, the secrecy gain is $\frac{4096}{301}$.

\end{document}